# The Alt-Right and Global Information Warfare

Bevensee, Emmi
PhD Student, School of Information, University of Arizona
Senior Fellow, Center for a Stateless Society
Tucson, Arizona, USA
emmibevensee@email.arizona.edu

Ross, Alexander Reid
Instructor, Department of Geography, Portland State University
Portland, Oregon, USA
aross@pdx.edu

**Abstract-** The Alt-Right is a neo-fascist white supremacist movement that is involved in violent extremism and shows signs of engagement in extensive disinformation campaigns. Using social media data mining, this study develops a deeper understanding of such targeted disinformation campaigns and the ways they spread. It also adds to the available literature on the endogenous and exogenous influences within the US far right, as well as motivating factors that drive disinformation campaigns, such as geopolitical strategy. This study is to be taken as a preliminary analysis to indicate future methods and follow-on research that will help develop an integrated approach to understanding the strategies and associations of the modern fascist movement.

## I.   Introduction

Fascism is both a palingenetic and syncretic form of ultra-nationalism that frequently co-occurs with traditionalist fundamentalism, ethno-supremacy, authoritarianism, anti-cosmopolitanism, and opposition to open markets and liberalism. Although white supremacist groups manifest as far-right movements opposed to freedom and equality, closer scrutiny tends to reveal overlap with a syncretic definition of fascist movements more broadly [11] [68] [77]. While fascist and white-supremacist dissemination methods have historically occurred through pamphleting, subcultural aesthetics, and word of mouth, it now happens at a much larger scale through the internet. Additionally, on the social media landscape state actors are able to engage in information war at much larger level than ever before including by promoting extremist movements within other countries. As a result of the monumental increase in dataset size, analysis of this problem requires the tools of data science in order to generate wider reaching analyses of both mainstream social media and more niche forums. Tools like network mapping and natural language processing research on these social networks such as twitter provide a blueprint for better understanding this re-emergent threat on an international scale.

## II.   Literature Review (*Fascist, Alt-Right, and Hate-group Specific Data Science*)

While historical scholars of fascism such as Hannah Arendt and the Frankfurt School provide insights into the psychology and context that gives rise to fascist movements, they require an update into the modern context. Therefore, it becomes important to present a review of available sources of information on modern fascism, and particularly the Alt-Right.

An important scholar on the phenomena of Alt-Right internet based radicalization is Matthew Lyons who authored, *Ctrl-Alt-Delete: The Origins of the Ideology of the Alternative Right* and works, in part, for the cutting edge think tank, Political Research Associates, alongside other scholars of modern white-supremacist and fascist movements such as Spencer Sunshine. There is little modern study of the Alt-Right specifically from the academic sphere and instead comes either directly from the mouth of babes such as "An Establishment Conservatives Guide to the Alt-Right" by Bokhari and Yiannopolous [15], or the infamous, "Normies Guide to the Alt-Right" by the neo-Nazi Andrew Anglin [5]. Beyond these self-proclamations, the most in-depth source of additional research comes from various antifascist publishers and authors such as Mark Bray and Shane Burley, as well as decentralized antifascist research groups such as Rose City Antifa, One People's Project, and NYC Antifa.

Although crude in many respects, one important study provided a kind of natural language processing (NLP) study of the fascist extremist board 8chan /pol/ that looked at language usage and exposed a wide range of credible threats to violence resulting from the networked support and vectors for radicalization of the gritty forum [49]. This mirrored the more scholarly findings of "Stormfront is Like a Second Home to Me" by De Koster and Houtman in 2008 [2], which utilized qualitative analysis to suggest that online forums help white supremacists feel less stigmatized while solidifying and radicalizing their ideology. Ross's recent guide, *Against The Fascist Creep*, exposits a broader strategic approach by fascists to "enter" political and social milieux in order to disseminate [11]. While such historical patterns still apply, the modern fascist movement has distinct characteristics and threats.

Although there is an abundance of historical study on hate groups, and even more so Islamic extremism, there is comparatively little actual data science applied to movements such as the Alt-Right and their more extreme descendents' mobilization through the internet. In "Hate Online: A Content Analysis of Online Extremist Internet Sites," Gerstenfeld, Grant, and Chiang [41] found that "the Internet may be an especially powerful tool for extremists as a means of reaching an international audience, recruiting members, linking diverse extremist groups, and allowing maximum image control." This is reinforced by "Cyberhate: The Globalization of Hate" where Perry and Olson [25] showed how the internet has created a medium by which historically fractured white nationalist and fascist movements are able to create a collective and supra-national identity thus strengthening the movement in turn. Through network analysis focusing on reciprocal following and interactions of prominent white supremacist twitter accounts, Callaghan et. al. [14] uncovered deeply internationalized relationships developed over topics underlying extreme right ideologies. In an innovative approach, Burris, Smith, and Strahm [82] studied external links of white supremacist forums and websites to analyze bridges and cleaves in the movement. They found that:
*Interorganizational links are stronger among groups with a special interest in mutual affirmation of their intellectual legitimacy (Holocaust revisionists) or cultural identity (racist skinheads) and weaker among groups that compete for members (political parties) or customers (commercial enterprises). The network is relatively isolated from both mainstream conservatives and other extremist groups. Christian Identity theology appears ineffective as a unifying creed of the movement, while Nazi sympathies are pervasive. Recruitment is facilitated by links between youth and adult organizations and by the propaganda efforts of more covertly racist groups. Links connect groups in many countries, suggesting the potential of the Internet to facilitate a white supremacist "cyber-community" that transcends regional and national boundaries.*
This research implies that while the internet may facilitate greater outreach and unifying collective identity, it also brings schisms to the surface contradictingly amplifying fractures as is seen in the constant infighting and purity spirals of forums like 8chan [49].

More recently, Shannon Jones [51] published, "Mapping Extremism: The Network Politics of the Far-Right" in which she used network and discourse analysis to find that ultra-nationalist parties utilized extra-parliamentary networks on social media to help facilitate and sustain political takeovers in Europe. Similarly Bevensee and Yershov [37] showed that fascist movements employ memetically dangerous rhetorical devices in order to build capacity in attempting violent acts as well as larger political coups. They further showed how there is an explicit and conscious attempt to fight the social stigmatization of fascism by widening the Overton window through such online and discursive strategies. An Italian-specific study by Parenti and Caini [69] showed how extreme-right groups were exploiting "the Internet for diffusing propaganda, promoting 'virtual communities' of debate, fundraising, and organising and mobilising political campaigns," thus confirming the link between online and offline actualization of extremist goals (273).

The vast majority of studies surrounding radicalization surround the topic Islamic terrorism, despite the fact that there the threat of violence from fascist, far-right, and white supremacist groups is statistically greater [83]. Snopes reporter Dan MacGuill [84] wrote that "Not a single death has resulted from terrorist activity by a Muslim extremist refugee." However he also cited the faulty 2017 study by the U.S. Government Accountability Office that oversimplified the Pulse nightclub shootings by Omar Mateen as Islamic extremism, which led them to conclude that far-right attacks were lower in number than those of Islamic extremism. A more detailed study of Mateen's relationship to Islam revealed that it was spotty at best, as both his Imam and ex-wife confirmed [47]. Alt-Right and Islamic extremist movements have also been shown to share similar values from different angles [59], evidenced most explicitly by the semi-ironic "white shariah" movement coming out of the Alt-Right [81]. Although there are many complexities to counting and defining what makes a "fascist" or "white supremacist" crime versus "Islamic extremism," there is nonetheless a clear rise in hate groups [40] with a concomitant rise in hate crimes [39].

The Southern Poverty Law Center study of Alt-Right-linked killings released in 2018 found over 100 people murdered with the majority of the killers being under 30 (thus very exposed to the internet) and all male [67]. This is not surprising when reflecting on the research done in "A Psychological Profile of the Alt-Right" which found that adherents "expressed higher Dark Triad traits, social dominance orientation, and authoritarianism; reported high levels of aggression; and exhibited extreme levels of overt intergroup bias, including blatant dehumanization of racial minorities [6]. This study is further reflected in the Associated Press call to encourage reporters to not call adherents by the specific term "Alt-Right," which they deemed to be a deception designed to hide their true allegiances to white supremacy movements [17]. Meanwhile the right claims the status of victim to Islamic, immigrant, and leftist violence despite responsibility for the clear majority of political violence [85]. Despite whatever semantic difficulties exist, the demand for broad-based data science on the topic of fascism and white supremacism, more broadly, and the Alt-Right, specifically, is greater than ever.

### A. Disinformation and Botnets

Peter Hernon [29] writes that "Inaccurate information might result from either a deliberate attempt to deceive or mislead (disinformation), or an honest mistake (misinformation)." The Alt-Right presents a complexification of this simplified notion in that individual actors may varyingly believe in the racist or fascist political propaganda they spread, or they may be intentionally using information misleadingly to achieve a political ends. Additionally, while they may be making an "honest mistake" personally, they can be unknowingly a part of a larger geopolitical struggle of disinformation through state-run media outlets pursuing a political agenda.

Another tool in the arsenal of state-based disinformation campaigns is social-media bots as part of the information warfare of hybrid-warfare [89]. Social-bots have begun to represent a large share of the social media landscape and while some are innocuous or even helpful, many can cause damage [74] either through things like automatic retweeting of disinformation or of politically motivated propaganda spread [44]. It is difficult, however, to identify malicious twitter accounts [35] or fake news (Shu, et. al. 2017). Kumar and Shah [32] have made significant progress in identifying the typologies of false information spread through social media and the various effectiveness of its many forms. Ferrara et al [74] also created a massive sting operation called "Bot or Not" that utilized both crowdsourced and machine learning algorithms to detect bots on social media through various factors. For the purposes of this paper, we utilize the language of disinformation more broadly to highlight the ways in which Alt-Right mobilization is rooted in faulty notions for the purposes of political power grabs that can manifest on the larger multi-national theatre of geopolitical warfare.

### B. Gaps in the Literature

It is clear that the interconnectivity and global reach of the internet has brought new opportunities and challenges to the field of politics and policy. The same fascist movements that see "globalism" as a threat to their integrity have since begun to rely and even excel at utilizing the internet as a means of borderless connectivity for the purposes of violently anti-multicultural mobilization. By analyzing Alt-Right movements using big data methodologies on social media, we can reveal the extent to which they have succeeded and how best to thwart their movements. To the extent that these movements are being steered by larger, state-based geopolitical projects, they will be more difficult to unravel. However, data science provides the tools needed to tease apart the connections underpinning these modern movements. This research is all the more important while many western powers over-focus on foreign Islamic threats rather than homegrown (if internationally manipulated) fascist threats veiled as the "Alt-Right".

### III. Historical and Present Context

Much of the Alt-Right's international outlook stems from international correspondence groups forged through affinities among fascists during the 1980s and '90s. Attempting to fuse revolutionary left-wing ideology with far-right ethnonationalism, groups like American Front entered into association with the Italian Terza Posizione (Third Way), France's Troisieme Voie (also Third Way), and the influential Belgian fascist, Jean-Francois Thiriart. Together, this so-called "International of Mailboxes" circulated group literature, manifestos, and strategic documents to set forth a strategy for international fascism. Amid the decline of the Soviet Union, most of these "Third Positionists" pursued a "National Bolshevik" ideology, which supported a kind of revolutionary, ethno-nationalist model of Soviet-like governance in Western Europe in favor of North Atlantic liberalism purportedly imposed by NATO [18].

In the early 1990s, Thiriart traveled to Russia to promote a kind of Eurasian integration under a federated system of strong, ultranationalist states, while his Italian disciple, Claudio Mutti, joined former Troisieme Voie organizer, Christian Bouchet, in advancing a geopolitical strategy promoted by Russian fascist, Alexander Dugin [18]. In Dugin's influential 1997 text, *Foundations of Geopolitics*, the National Bolshevik explicitly developed the strategy for intriguing within the social and political dynamics of the North Atlantic in order to bring about the disintegration of liberalism [36]. Utilizing the geopolitical notions of the Eurasian "Heartland" and Atlantic "Rimland" initially conceived by British geographer, Halford Mackinder, Dugin layered fascist myths of the Aryan "Sonnenmensch" over an idea of a united Eurasia to promote a global Russian empire from Dublin to Vladivostok and southerly to the Indian Ocean.

*"In this respect, Heartland's position is clear: it is necessary to resist the USA's Atlantic geopolitics, on every level and in every region of the world, in an attempt to fully weaken, demoralise, deceive and ultimately defeat the opponent. It is particularly important, moreover, to introduce geopolitical disorder into domestic American reality, encouraging all kinds of separatism, a variety of ethnic, social and racial conflicts, actively supporting all dissident movements: extremists, racists and sectarian groupings, destabilising the domestic process in the USA. At the same, this means supporting isolationist tendencies in American politics and the arguments of (often right-wing Republican) circles that believe the USA should confine itself to its domestic problems."* [36]

While Dugin's text found wide currency and influence among high levels of the Russian state, including the General Staff of the Armed Forces, the Putin regime installed in 2000 pursued a foreign policy largely amenable to Western interests until Western support for Kosovo's independence in 2007 [19]. In 2008, Russia invaded Georgia after a brief period of escalation involving the early model of modern cyber attacks seen currently in Ukraine. As competition over influence and control over resources in Central Asia ramped up in the wake of the 2008 financial crisis, relations between the Kremlin and the West grew increasingly tense as Putin blamed the decadent Western liberals for mass public protests on the heels of his 2011 re-election [75]. Russian state media turned the anti-Western sentiment toward support for protest movements in the US like Occupy Wall Street, while downplaying dissent within Russia [22].

Russian media became a hub for right and left-wing political activists to express their dissent against the US, and particularly against liberalism. The Alt-Right fit into this strategy, having emerged in 2008 largely through the influence of "anti-interventionist" or "isolationist" far-right ideologues in college organizing—particularly in relation to Richard Spencer's reports on the Duke Lacrosse Scandal, as well as the emergence of both the Taft Society and the Youth for Western Civilization [80] [70] [3]. Hence, Russian state-sponsored media followed Dugin's strategy of pitting left and "isolationist" far-right against the liberal center as Putin increasingly gravitated toward a foreign policy focused on Eurasian integration.

In 2013, Russia launched Sputnik, which advanced more radical perspectives from the left and right, often in collaboration. While Russian media corresponded with the likes of Mutti and Bouchet, the Alt-Right deepened its contacts with Dugin through groups like the Traditional Workers Party created by former Youth for Western Civilization organizer Matthew Heimbach. Meanwhile, conspiracy theorist Alex Jones hosted Dugin on his Infowars YouTube channel, and made a guest appearance on Dugin's far-right Tsargrad television channel. The integration between Alt-Right and the international fascist movement, as located through the work of Thiriart's cohort in the 1980s, was furthered by the trans-national development of Identity Evropa, a group that spun off from Youth for Western Civilization, borrowing its ideology from the Identitarian ideology promulgated by French fascist, Guillaume Faye, whose "Euro-Siberianism" is approximate to Dugin's Eurasianism [61]. Thus, Russian media, Duginism, and the European fascist movement, more broadly, played an important role in the maturation of the Alt-Right from its inception in 2008 up to its mainstreaming during the Trump campaign.

## IV. Methodology

Our research uses a combination of data mining techniques to unearth networks and implications of news providence through the study of links [82]. For the initial round of investigatory testing we utilized around 50 twitter accounts from prominent "Alt-Right" and "alt-light" accounts. The alt-light is a variation of the Alt-Right that splintered from the plainly fascist elements during the Trump campaign, while continuing to utilize obfuscation and conspiracy theory to find entryways into the mainstream. We utilized several Python packages, including Tweepy, to gather around 5000 tweets from 20 Alt-Right and 20 alt-light accounts, and then extracted and expanded the links from within those tweets. We then counted the frequency of unique base URLs from both the Alt-Right and alt-light accounts in order to compare them to each other.

## V. Initial Findings

External link analysis of the initial data revealed that the majority of both Alt-Right and alt-light accounts rely on other social media sites and accounts for information sources, rather than credible news outlets or journals. Aside from retweets, which were the majority of links, people frequently linked to things like Facebook posts, instagram accounts, and to their own personal accounts on gab.ai, which is a "free speech" social media platform that has come under fire for allowing the incitement of violence and its ties to prominent neo-nazis [52]. Accounts cite YouTube videos more often than anything else, often linking to their own YouTube channels as seen by YouTube micro-celebrities such as Lauren Southern and Bridget Pettibone, who both recently met with Dugin in Russia [60]. Echoing the results of research into Islamic terrorism, we found that YouTube is a vector for white supremacist radicalization [21].

Similarly, the accounts we studied often link to either their own personal blogs or the blogs with which they work. This is illustrated by figures like David Duke and Peter Sweden creating personal websites, and by regular retweeting of white supremacist "alternative news" or research outlets such as theoccidentalobserver.net, nationalpolicy.institute, and VDare. Among the most explicit iterations to show up are "Fash the Nation" which is a project of "The Right Stuff" and hosts podcasts such as "The Daily Shoah." As may be obvious from the thinly veiled anti-Semitism and fascist allusion in the titles, this is an extreme Alt-Right podcast. They are responsible for the propagation of the anti-Semitic memes, like the "(((echoes)))" phenomena used to identify supposed widespread Jewish control [30]. They came under fire from other Alt-Right organizations, however, when their founder was doxxed and revealed to have been married to a Jewish woman [54]. Additionally, "The Political Cesspool" appeared in the Alt-Right links. A white-supremacist podcast hosted by James Edwards, who wrote a book called, "Racism,

Schmacism" and is on the board for both the American Freedom Party and the Council of Conservative Citizens, "Political Cesspool" is a clear example of fascist media efforts online, as well as deep coordination between fascist groups that attempt to disguise themselves as radical-right conservatives (e.g., the "Freedom Party" identity popularized by the Austrian Freedom Party, or FPO) [13].

Counter to these and other explicit white-supremacist sites, there was a frequency of Jewish or Israeli sites linked. These were often either making fun of their panic in sites like gab.ai or attempted "gotchas" at perceived loopholes in their political views. This shows that, in addition to their own alternative news ecosystem, the Alt-Right also engage in a degree of monitoring of outlets that they perceive as ideological enemies.

Many accounts seek to build credibility by citing multiple blogs which, upon analyzing WhoIs registry data, are revealed to be controlled by the same organization. The sites theamericanfreedomparty.us, conservative-headlines.org, and thedailytrump.org are all owned by the Nevada-based American Freedom Party which stems from the so-called "Third Positionist" (fascist) political tendency [13]. This trend of "refraction," through which a single message is reproduced through different "wrappers" or political perspectives to give the perception of the agreement of a heterogeneous plurality has been observed in conspiracy theory networks supporting Russian geopolitical imperatives in Syria [44].

The import of personal blogs is made all the more clear by the degree to which the aggregation of websites only shared once or twice appeared in both the Alt-Right and and alt-light twitter accounts. This reveals a skew towards personal blogs such as national-socialist-worldview.blogspot.com and refugeesettlementwatch.wordpress.com and away from larger media outlets with track records of credibility and fact checking.  For example, Refugee Resettlement Watch is run by Ann Corcoran who frequently posts on the white supremacist blog American Renaissance, which was founded by the white nationalist, Jared Taylor, to "prove" the inferiority of black people to whites and has become a common "source" among racist groups [50]. In addition to these "alternative media" hubs, there is also the presence of a number of nationalist European or pseudo-Eurasian news blogs such as traditioeuropae.wordpress.com, breizatao.com, europa-terra-nostra.com, or orientalreview.org.

Aside from this highly dispersed and diffuse, alternative news ecosystem of personal blogs and YouTube videos, there is also a density of links to Kremlin media outlets such as Russia Today (RT) and Sputnik News. The ability of Russian state-controlled media outlets to drive their narratives through prominent Alt-Right figureheads aligns with the historical strategy of people like Dugin and his affiliated Izborsky Club, as well as the modern tenants of "hybrid war" (what Dugin calls "fourth political warfare") often associated with the "Gerasimov Doctrine" of the leading officer in the General Staff of the Russian Armed Forces. Despite featuring prominently in the Alt-Right links, these Russian outlets did not show up amongst the alt-light tweeters. This shows a degree of nativism but may not indicate a lesser extent of syncretic strategy, insofar as the alt-light is more prone to link to domestic, ostensibly-conservative sites and less likely to link to foreign media, but does not have qualms with Tucker Carlson inviting left-wing guests on to his FOX News TV show. Carlson's The Daily Caller, which, despite being more mainstream than many of the other sources, has engaged in various forms of conspiracy pandering such climate change denialism, also appeared in the Alt-Right links [27]. The Daily Caller has gained prominence on the far right by taking radical and sometimes violent positions, for instance encouraging people to use their cars to run-over protestors, as happened in the murder of Heather Hayer during the Charlottesville "Unite the Right" rally [28].

There was one English speaking reference from the Alt-Right to Dugin's Russian blog 4pt.su by Arktos Media, mourning it's temporary or permanent outage. Arktos is an Alt-Right media outlet that translates Dugin alongside the fascist French founder of the Nouvelle Droite and pivotal inspiration for the Alt-Right, Alain de Benoist, who also spoke at a conference hosted by Richard Spencer's Alt-Right and white supremacist think-tank National Policy Institute [66].

Breitbart featured prominently in the alt-light accounts but about half as often in among the Alt-Right twitterers. The former head of Breitbart, Steve Bannon, was also the former chief strategist to Donald Trump. Bannon has shown strong ideological ties with the Eurasianism of Dugin and Putin himself [45]. This seems strange, considering the strong historical relationship between the Alt-Right and Breitbart, a la figures like Milo Yiannopoulos [15] who helped smuggle white supremacy into the mainstream and ultimately into the White House. Yet it is still important to note that Breitbart remains closer to an edgy conservative outlet of the "radical right" ilk, rather than being purely Alt-Right. Sites like these make widening the Overton Window more possible as indicated by Bernstein's famous BuzzFeed article on the subject [42].

Among the sites most implicated in both Russian disinformation and syncretic conspiracy theories are Veterans Today, 21st Century Wire, ZeroHedge, GlobalResearch.ca, MintPressNews, and The Duran [44]. We analyzed accounts associated with these sites separately from the Alt-Right and alt-light sites, and cross-referenced them for correlations. Only ZeroHedge appeared in our initial data collection and only slightly in alt-light. It is possible that the higher level figures we examined rely less on direct references to these sites than their followers, in addition to the fact that we are not yet studying the types of bots implicated in propagating conspiracy theories at the behest of state actors. However, ZeroHedge made another appearance in a Latent Semantic Indexing of 500 tweets between Alt-Right handles and their followers. Although inconclusive, this does suggest that an analysis that expands the handles beyond these prominent figures would feature such sites more prominently. In the sphere of conspiracy theories, however, Infowars was found in both Alt-Right and alt-light link counts, as well as other websites implicated in conspiracy theory propagation, such as thefreethoughtproject.com [38].

The most prominent and clearest connection between the Alt-Right and conspiracy theory sites in our conspiracy theory selection is called The Unz Review, which appeared very frequently amongst the Alt-Right twitter handles. The Unz Review is a "mix of far-right and far-left anti-Semitic crackpottery, from 9/11 'truther' and conspiracy theorist Paul Craig Roberts to 'Holocaust industry' critic Norman Finkelstein, who believes Jews exploit the Holocaust to justify oppressing Palestinians" [87]. This website represents the heart of Alt-Right disinformation landscape, in that it sees itself as a radical opposition to the mainstream that transcends the traditional left-right political binary in order to propagate white-supremacist narratives amidst a wide range of conspiracy and propaganda.

Among the sites that support the Unz Review is the anti-Semitic and pro-Kremlin website russia-insider.com [86], which also featured in the Alt-Right websites linked. Russia-Insider keeps a running feed of articles posted on The Unz Review. Russia-Insider was founded in tandem with a larger effort to generate a more positive view of the Kremlin in the US by taking a critical approach to US politics from a Russian perspective while advancing pro-Russian conspiracy theories. Russia-Insider's top sources of traffic, according to Amazon's Alexa site overlap tool are searches for ZeroHedge, and the top sites visited before it were Sputnik News and RT.

Aside from the linkages fostered by hyperlinks and audience clustering, as rendered through site metrics, it is important to observe direct, personal linkages in this alternative newscape. Russia Insider was co-founded by a Gilbert Doctorow, a columnist for the alternative news site Consortium News, which publishes conspiracy theorist Caitlin Johnstone, who openly supports an alliance between left and far-right forces against the liberal center. Doctorow also co-founded the pro-Kremlin interest group "American Committee for East-West Accord" (ACEWA) in 2014 via an event in Moscow known as the Russia Forum, which is convened by the American University in Moscow headed by long-time far-right operator, Edward Lozansky. Lozansky's University hosts a "think tank" that includes Doctorow, Duginists like Mark Sleboda, political activists like the Ron Paul Institute's Daniel McAdams, and Unz Review columnist, Anatoly Karlin.

Present at the "round table" that conceived the ACEWA was a Duginist named Andrew Korybko, whose analyses have been published by the Russian Institute for Strategic Studies, a think tank noted for its involvement in strategizing the 2016 elections meddling [57]. The ACEWA was subsequently launched in Brussels by Doctorow with another "round table" that included a far-right, pro-Kremlin associate of Dugin named Aymeric Chauprade, suggesting an inclination to draw together left and right with the intention of supporting relations with the Kremlin. Doctorow has also written in support of Dugin's associate, Mateusz Piskorski. With Doctorow, the ACEWA was co-founded by a contributing editor to The Nation, Stephen F. Cohen, and its website editor is American Conservative contributor, James Carden [11].

The convergence of these interests—specifically, Doctorow's engagements with Consortium News and Russia Insider, Cohen's association with The Nation, and Carden's involvement with The American Conservative—indicates a widespread willingness to engage in a far-reaching effort uniting ideological positions in favor of a "new detente" with the Kremlin, vis-a-vis a coalition spanning far right and radical left. This engagement mirrors Dugin's strategic proposition that the Kremlin should propagate and spawn both extreme right and left wing politics as a means of destabilizing the US while also increasing the ground upon which larger, syncretic fascist projects, such as the Alt-Right, can be sewn [12].

Further, Dugin has shown incredible support for both Donald Trump [73] and the Alt-Right. Evidence of the impact of Duginist political philosophy on concrete Russian policy can be seen in the fallout surrounding the expansive involvement of the Russian "Internet Research Agency" on the 2016 election of Donald Trump [79]. Through an elaborate troll farm, they were able to create thousands of bot accounts that promoted right and left-wing opposition to liberalism while propagating, and even funding, various political events, including protests. It might be said that the general idea of tampering in foreign elections through various forms of media is well-worn strategy practiced by different countries and corporate actors, including the US; however, Russian media strategy reveals a more sophisticated, internet-age approach to geopolitical manipulation that, if successful, could support internet-led fascist mobilizations in the US and elsewhere. The work of Kate Starbird [44] and others on disinformation and conspiracy theory networks on Twitter has begun to unravel the ways in which movements like the Alt-Right can create an

"alternative media" bubble [1] that is vulnerable to state interventions. Further study of things like botnets, as well as news sources of the Alt-Right through things like external link analysis, could begin to unravel the extent to which these movements are a part of a larger geopolitical project.

## VI. Need for further research

This initial dataset was intentionally constrained to ensure methodological consistency and code capacity in order to scale analysis later. As a follow up we will create a metric for "friendship" that can be used to generate a much more massive dataset following the other similar research procedures. From this big data set we will create network maps in addition to more thorough analysis of external links. We will then also scale external link analysis in order to create network maps of websites to investigate how often they reference each other beyond just twitter. From this position we could also look at server IP addresses to determine a rough analysis of where different sites are being hosted.

As we scale the number of accounts we can also begin to incorporate methods that allow for the inclusion of potential socialbots and more fringe accounts. This will likely reveal more clearly the relationship of the Alt-Right to conspiracy theories and "alternative news" sites.

## VII. Conclusion

Amidst the resurgence of fascism and authoritarianism in popular culture, there is a concurrent global information war afoot. State super-powers are warring through manipulating the information sources and social media discourses of different political groups in foreign nations. Our initial data suggests that the Alt-Right leans heavily on Kremlin controlled media sources as a critical backbone of their ideology, thus playing a role in their concomitant violence. Further scaling of data mining analysis of Alt-Right twitter networks will give greater insight into the shape and contour of fascist disinformation and influence networks in the age of big data. Undermining the instrumental rationality of authoritarianism through data analysis gives us the opportunity to push-back on the empowerment of proto-fascist movements across the globe.